\documentclass{article}
\newcommand{\be}{\begin{equation}}
\newcommand{\ee}{\end{equation}}
\def\bea{\begin{eqnarray}}
\def\eea{\end{eqnarray}}
\begin{document}

\title{A Note on Strings in a Rindler Background\\[1cm]}

\author{Sudarshan Ananth\thanks{ananth@phys.uf\,l.edu}\\[0.5cm]
{\it {Institute for Fundamental Theory}}\\
{\it {Department of Physics, University of Florida,}}\\
{\it {Gainesville, FL 32611, USA}}}
\maketitle

\begin{abstract}
{\noindent}We study bosonic strings in a Rindler background using a D-23 brane. This model is shown to be directly related to the orbifold approach to Rindler strings. We propose a duality between these pictures based on comparisons of their closed string spectra.
\end{abstract}

\section{Introduction}

The problem of Strings in a Rindler background is an important one and remains intractable for arbitrary temperatures. The near horizon geometry of most black holes is well approximated by Rindler and a better understanding of strings in this background could serve as a handle to deal with more complicated geometries like Schwarzschild. The familiar connection between the Rindler vacuum (at temperature ${\Theta}={1\over 2{\pi}}$) and the Minkowski vacuum corresponds to the unique map that takes the non-linear Rindler sigma model into the linear Minkowski sigma model. Any change in temperature away from this value causes the geometry to become conical with a curvature singularity at the origin. This problem may be overcome to some extent by viewing the geometry as an orbifold of ${{R_2}\over Z_N}$~\cite{dab2}. This reproduces the conical geometry and allows us to calculate at temperatures that are integer multiples of $\Theta$. The purpose of this note is to use D-branes to reproduce these same temperatures. It appears as if the orbifold and D-brane pictures are very similar. Further we claim that the closed string spectra in these theories are the same.

\section{Strings in Rindler}

\subsection{Rindler Space}

The Rindler `wedge' is Minkowski space mod $Z_4$. This space (of uniformly accelerated observers) has horizons and is a good toy model for black hole physics. The metric for 2-D Rindler space is obtained from the Minkowski metric,

\be
\label{mm}
d{s^2}=d{t^2}-d{x^2}
\ee
by the following transformations,

\bea
\label{mtor}
t={1\over a}{e^{a{\xi}}{\sinh{a\eta}}}\nonumber \\
x={1\over a}{e^{a{\xi}}{\cosh{a\eta}}}
\eea
which yield,

\be
\label{rindm}
d{s^2}=e^{2a{\xi}}(d{{\eta}^2}-d{{\xi}^2})
\ee

Black Hole entropy is an interesting quantity and its calculation requires a knowledge of the partition function. This is computed using the Euclidean continuation of the Rindler metric,

\be
d{s^2}={{\rho}^2}d{{\theta}^2}+d{{\rho}^2}
\ee
where ${\theta}=i{\eta}$ is the periodically identified time coordinate with period ${\beta}$ (the inverse temperature) and ${\rho}=e^{a{\xi}}$. Any change in $\beta$ away from $2{\pi}$ causes the Euclidean geometry to become conical with a curvature singularity at the origin.

\subsection{The Rindler Sigma Model}

A string sweeps out a two dimensional worldsheet which is described by two parameters, ${\sigma}$ and $\tau$ (proper time). The Nambu-Goto action for a string propagating in space time may be rewritten in a more convenient form, introduced by Brink et al~\cite{brink},
\be
\label{five}
S_P=-{1\over {4{\pi}{\alpha^{\prime}}}}{\int}d{\tau}d{\sigma}{(-{\gamma})}^{1\over 2}{\gamma^{ab}}{\partial_a}{X^{\mu}}{\partial_b}{X^{\nu}}{{\eta}_{\mu\nu}}
\ee
where ${1\over {2{\pi}{\alpha^{\prime}}}}$ is the string tension, ${\gamma^{ab}}$ is the worldsheet metric, $X^{\mu}$ are scalar fields (from a worldsheet point of view) and ${{\eta}_{\mu\nu}}$ is the spacetime Minkowski metric. With the choice of a Euclidean world sheet metric ${\gamma^{ab}}={\delta^{ab}}$, the equations of motion may be solved to obtain a very simple solution for the closed string fields $X$,
\be
X=x+p{\tau}+{i\over 2}{\sum_{n{\neq}0}}{{\alpha_n}\over n}e^{-2in({\tau}-{\sigma})}+{i\over 2}{\sum_{n{\neq}0}}{{{\tilde {\alpha}}_n}\over n}e^{-2in({\tau}+{\sigma})}
\ee
with $x$ and $p$ being the center of mass position and momentum. $\alpha$ and ${\tilde {\alpha}}$ are the left and right moving oscillators respectively.
This model is easily generalized to curved spacetimes by replacing the flat metric ${{\eta}_{\mu\nu}}$ in (\ref {five}) by the general metric ${G_{\mu\nu}}$.
\be
S_P=-{1\over {4{\pi}{\alpha^{\prime}}}}{\int}d{\tau}d{\sigma}{(-{\gamma})}^{1\over 2}{\gamma^{ab}}{\partial_a}{X^{\mu}}{\partial_b}{X^{\nu}}{G_{\mu\nu}}
\ee
This is where the problem arises because the Rindler metric causes this theory to become coupled due to the factor of ${e^{2a{\xi}}}$ in (\ref {rindm}). As long as this space remains flat (${\Theta}={1\over {2\pi}}$), the inverse transformations to (\ref {mtor}) allow us to map this non-linear sigma model back to the free sigma model. This is not possible at other temperatures.

\subsection{${{R_2}\over {Z_N}}$ Orbifolds}

Strings do not propagate on a cone but as mentioned earlier the geometry may be described as an orbifold of ${{R_2}\over Z_N}$(for integer $N$). An orbifold identifies spacetime points under a discrete symmetry. In the case of ${{R_2}\over {Z_N}}$, we define the operator $\alpha$,

\be
\label{orb}
{\alpha}:X=e^{{2{\pi}ik}\over N}X
\ee
which identifies various $X$ that are related to each other by the action of $\alpha$. These identifications cause the manifold to become conical with a singularity at the origin. $N$ determines the periodicity $\beta$ (and the deficit angle) and hence controls the temperature of this space. $k=1{\ldots}N-1$ correspond to the twisted sectors in the spectrum. The orbifold group acts on the string field $X$ by multiplication. The partition function is computed as a trace weighted by the exponential of the Hamiltonian and momentum with the projection operator inserted within the trace~\cite{pol}. For example, in the case $N=3$,
\be
Z={(q{\bar{q}})}^{1\over 48}Tr({{1+{\alpha}+{\alpha^2}}\over 3}{q^{L_0}}{\bar{q}}^{{\tilde L}_0})
\ee
where $q=e^{2{\pi}i({\sigma}+i{\tau})}$, $L_0$ is the zero mode of the Virasoro generators and ${\alpha^3}=1$.

The drawback in this procedure being that the range of accessible temperatures is limited by the integer nature of $N$, to be of the form ${N\over {2{\pi}}}$.

\section{D-23 Branes and the Rindler Metric}

We work in a 26 dimensional spacetime of the form ${M_{24}}{\times}{E_2}$ where $M$ refers to a 24 dimensional Minkowski surface (D-23 brane) and $E$ a two dimensional Euclidean `transverse' space. $M$ and $N$ are spacetime indices and run over $0{\ldots}25$, $\mu$ and $\nu$ are brane indices and run from $0{\ldots}23$ and $a$ and $b$ run over 1, 2 (transverse indices). We fix a D-23 brane at $x^a=0$. The action for this D-23 Brane reads~\cite{cikl},

\be
\label{act1}
S=-{\it f}{{\int}_{\Sigma}}{d^{24}x}{\sqrt{-\widehat{G}}}+{{M_P}^{24}}{\int}{d^{26}x}{\sqrt{-G}}\;R
\ee
where $G$ is the spacetime metric with Ricci scalar $R$. ${\it f}$ is the D-brane tension, ${M_P}$ the 26 dimensional Planck mass and ${\Sigma}$ is the surface defined by $x^a=0$. $\widehat{G}$ is the metric induced on the 23-brane,

\be
\label{indm}
{{\widehat{G}}_{\mu\nu}}={{\delta_{\mu}}^M}{{\delta_{\nu}}^N}{G_{MN}}{|_{\Sigma}}
\ee
The equations of motion for (\ref {act1}) are

\be
\label{eom}
R_{MN}-{1\over 2}{G_{MN}}\;R+{1\over 2}{{\sqrt{-\widehat{G}}}\over {\sqrt{-G}}}{{\delta_{M}}^{\mu}}{{\delta_{N}}^{\nu}}{{\widehat{G}}_{\mu\nu}}{{{\it f}\over {{M_P}^{24}}}}{\delta^{(2)}}(x^a)=0
\ee
Choose an Ansatz of the form,

\be
\label{ans}
d{s^2}={\eta_{\mu\nu}}d{x^{\mu}}d{x^{\nu}}+{e^{2{\gamma(x^a)}}}[d{{\rho}^2}+{{\rho}^2}d{{\phi}^2}]
\ee
for which

\be
R_{\mu\nu}=R_{{\mu}a}=0\;,\;{R_{ab}}={1\over 2}{G_{ab}}\;{R}
\ee
and the equations of motion yield,

\be
\label{omeg}
{\sqrt{G}}\;{R}={4{\pi}T}\;{\delta^{(2)}}(x^a)
\ee
where ${R}$, ${R_{ab}}$ are obtained from the two dimensional metric and we have defined the dimensionless quantity,

\be
T={{\it f}\over {4{\pi}{{M_P}^{24}}}}
\ee
Since for a conformally flat metric ${\sqrt{G}}\;{{R}}=-2{\partial^a}{\partial_a}{\gamma}$, our equation becomes

\be
{\partial^a}{\partial_a}{\gamma}=-{2{\pi}T}\;{\delta^{(2)}}(x^a)
\ee
from which we get,

\be
\label{ome}
{\gamma}(x^a)=-{1\over 2}{T}\;{\ln({{x^2}\over {q^2}})}
\ee
where $q$ is a constant of integration.

Assume ${T}<1$, substitute $\gamma$ back in the Ansatz and define a new coordinate

\be
r=e^{-{\beta}}{q^{(T)}}{{\rho}^{(1-T)}}
\ee
where $e^{-{\beta}}=1-T$.

The spacetime metric now reads

\be
d{s^2}={\eta_{\mu\nu}}d{x^{\mu}}d{x^{\nu}}+{d{r^2}}+{e^{-2{\beta}}}{r^2}d{\phi^2}
\ee
Thus our 26 dimensional space-time is now the product of a 24 dimensional Minkowski space and a 2 dimensional wedge, with deficit angle

\be
\label{rindt}
{\delta}={2{\pi}}[1-e^{-{\beta}}]={2{\pi}}\,T
\ee
The transverse space is a Euclideanized version of Rindler space and the D-brane tension serves as a thermostat for this space.

\section{Brane-Orbifold Duality}

Our analysis so far has been very general (holds for arbitrary dimensions) and is not directly related with strings. We want to use this constructed background to study bosonic closed strings. Our choice of a 23-brane (for 2-D Rindler) is explained by the fact that bosonic string theory is consistent only in 26 dimensions. The geometry of this D-23 brane model puts spacetime restrictions on fields in the transverse space. A choice of 
\be
\label{connect}
T=1-{1\over N} \;\;{\Rightarrow}\;\;{\delta}= 2{\pi}(1-{1\over N})
\ee
with integer $N$, demands that the closed string fields within the wedge satisfy,

\be
\label{com}
X=e^{{2{\pi}i}\over N}X
\ee
which is the same demand made by the ${{R_2}\over {Z_N}}$ orbifold. The partition function is computed in exactly the same fashion as the orbifold~\cite{dab2,stro}. The worldsheet at one loop is a torus. The partition function in the transverse space is simply the path integral of a single complex boson subject to twisted boundary conditions.
It follows then that the closed string spectrum in ${{R_2}\over {Z_N}}$ is identical to the closed string spectrum in the transverse space of our D-23 Brane (An orbifold has multiple `copies' of a string which arise from repeated identifications using (\ref {orb}) but this is true with (\ref {com}) as well. Further the important space is the wedge, the only space that really remains. The string itself has no way of knowing what causes this identification, so the manner in which the space is constructed is immaterial).
The D-brane approach may in fact be a more fundamental way of imposing the restriction (\ref{com}).

\section{Unresolved Issues}

At this stage, the brane tension appears as a free parameter. At specific values of this parameter, we have identified the connection with orbifolds. However the significance of these specific values remains unclear. We also need to understand why 26 spacetime dimensions are special from a D-brane point of view. The energy density or tension of the D-brane needs to be physically controlled. We suggest a couple of methods to achieve this tuning and hope to return to these issues later.

\subsection{Tachyon Potential}

The tuning process is tricky because the very mechanism of tuning the D-brane tension may interfere with the set up and mess up the geometry.
One method advocated by A. Sen~\cite{sen,sen2} is to use the tachyon potential to tune the D-brane tension. D-branes are usually described by boundary conditions on the open strings that end on them. However they may also be regarded as open string field theory solitons. Each D-brane in bosonic string theory has a tachyonic mode. It is believed that as the tachyon field $A$ rolls to the minimum of the tachyon potential $V(A)$, the negative contribution from the tachyon potential exactly cancels the tension of the D-brane. Since the brane itself disappears, this vacuum cannot support open string excitations. It has been shown that pushing the tachyon toward the side of the maximum where there is a local minimum corresponds to adding a positive definite boundary term to the world sheet action. The boundary state description associated with this `rolling' tachyon solution allows us to compute the tachyon potential energy contribution to the total energy density on the brane. For small displacements ${\tilde {\lambda}}$ of the tachyon field, the tachyon potential energy is $-{{{\tilde {\lambda}}^2}\over {2{g^2}}}$ where $g$ is the open string coupling and this adds to the initial energy density on the D-brane, effectively tuning it. For instance, in order to reproduce the wedge arising from an orbifold of ${{R_2}\over {Z_3}}$ we need a D-brane with,

\be
T={2\over 3} \;\;{\Rightarrow}\;\;{\delta}={{4{\pi}}\over 3}
\ee
which corresponds to $N=3$ in (\ref {connect}). A choice of ${\tilde {\lambda}}$ such that $({{1+cos(2\pi{\tilde {\lambda}})}\over 2})=2/3$, will yield a system with energy density that is 2/3 times the tension of the original D-brane. Note however that this is not a static system, and will evolve with time. In the classical approximation the energy density will remain constant at 2/3 times the tension of the original brane, but the other components of the energy-momentum tensor (pressure) will go to zero exponentially\footnote{Dr. Ashoke Sen, private communication}.

\subsection{External Fields}

The tension may also be tuned by turning on external fields (like a background B field). The effect of such a $B$ field is among other things to change the effective tension of open strings(stretched in the direction of the field). The value of the bosonic D p-brane tension is related to ${\alpha^{\prime}}$ by ${{\tau_p}^2}\;{\sim}\;{{\alpha^{\prime}}^{11-p}}$. For a background electric field alone, the effective tension for an open string is given by~\cite{sse},

\be
\label{gold}
{1\over {2{\pi}{{{\alpha}^{\prime}}_{eff}}}}={{1-{{\tilde E}^2}}\over {2{\pi}{\alpha^{\prime}}}}
\ee
The net brane tension will then have a contribution from strings lying along the field. These strings with modified ${\alpha^{\prime}}$  will alter the tension of the brane. However an external field may warp the cone (the $\beta$ functions must be computed in order to determine the exact effect on the geometry).

\section{Conclusions}
The main advantage of this analysis is the possible `extra' freedom afforded by the tension. Such a continuous parameter is unavailable in the orbifold approach and is a handicap for tuning the temperature. However we have illustrated the strong similarities in these approaches. Only certain values of the tension seem related to the orbifold case. At those values of the D-brane tension, where the connection with orbifolds is made, strings will propagate in this space. It is still unclear if strings propagate in this space at other values of the tension (which is what will provide us with a continuous range of temperatures). It will be interesting to understand what makes these values of the tension special. Tuning via the tachyon (or the background field) may provide clues as to what picks out these values.\\[0.5cm]

{\bf {Acknowledgements}}\\[0.5cm]
I would like to thank Dr. Pierre Ramond for his constant interest, support and guidance during the course of this work. I would also like to thank Dr. Zongan Qiu for many helpful discussions.

\end{document}